\documentclass[12pt,preprint]{aastex}
\def\kms{{\rm km\,s^{-1}}}

\def\au{{\rm AU}}
\def\lim{{\rm lim}}

\def\tot{{\rm tot}}
\def\bin{{\rm bin}}
\def\init{{\rm init}}
\def\cl{{\rm cl}}
\def\bh{{\rm bh}}
\def\circ{{\rm circ}}
\begin{document}

\title{Sgr A* Companion S0-2: A Probe of Very High-Mass Star Formation}

\author 
{Andrew Gould}
\affil{Ohio State University, Department of Astronomy, Columbus, OH
43210-1173}
\email{gould@astronomy.ohio-state.edu} 
\and
\author 
{Alice C.\ Quillen}
\affil{University of Rochester, Department of Astronomy, Rochester, NY}
\email{aquillen@pas.rochester.edu} 

\singlespace

\begin{abstract}
The star S0-2, which is orbiting Sgr A* with a 15-yr period, almost certainly
did not form in situ.  We propose that it was injected into this
close orbit by the tidal disruption of a massive-star binary, whose primary
was more massive than S0-2 and at least $60\,M_\odot$.  
From numerical integrations we find  
that 1-2\% of incoming binaries with closest approach equal to 130AU
leave the secondary in an orbit with eccentricity within 0.01  of that of SO-2.
If additional stars are found orbiting Sgr A*
with relatively short periods, they could be used to probe the formation
of massive stars in the Galactocentric region, even though the massive
stars themselves have long since perished.

\end{abstract}
\clearpage
 
\section{Introduction
\label{sec:intro}}

Almost a decade of high resolution infrared observations of the Galactic
center have revealed at least three stars that show substantial
proper-motion acceleration due to the supermassive black hole (BH)
associated with Sgr A* \citep{eckgen,ghez98,schodel,ghez03}.  
The most striking of these is S0-2,
which has a peribothron of only $q=125\,\au$ and a period of only $P=15\,$yr.
S0-2 has a number of important potential applications.  For example,
by combining radial-velocity and astrometric observations of S0-2
one could measure the mass, the distance, and the three
components of motion of Sgr A* \citep{visbin}, 
which is believed to lie almost exactly at the Galactic center.

However, the most
remarkable characteristic of the Sgr A*/S0-2 ``visual binary'' is
that it exists at all \citep{ghez03}.  S0-2 cannot have formed in place:
the gas cloud from which it would have condensed could never have
been larger than $\sim (m_1/M_\odot)^{1/3}\,\au$ or it would have been
tidally disrupted.  Here $m_1$ is the mass of S0-2.  Hence, it must have
formed at a much larger distance from Sgr A* and was either
originally on a highly eccentric orbit that took it within $\sim 125\,\au$
of Sgr A* or was perturbed onto such an orbit.  Then, it would have had to 
have been 
perturbed again at peribothron to bring its eccentricity from $(1-e)\ll 1$
to $(1-e)\sim 0.13$.  It is this last step that is problematic.

Spectra at 2$\mu$m taken by \citet{ghez03} exhibit Br$\gamma$ 
and HeI absorption lines and lack a CO bandhead.  These features suggest 
that S0-2 is
a O8-B0 dwarf star and so is massive, $m_1\sim 15\,M_\odot$, and young,
$\la 10\,$Myr.   The youth of the star puts a strong limit on the length
of time available to scatter the star into its current orbit since formation.
The dynamical friction time for such
a star in an orbit of semi-major axis $a\sim 10^4\,\au$ where it might
plausibly form, would be $t_{\rm df}\sim 300\,{\rm Myr}\,(a/10^4\,\au)^{-1/4}$,
even assuming that the stellar profile continued as a power law down to 
very small radii, as suggested may be the case by
\citet{alexander}.  Since several dynamical friction times would
be required to achieve the orbit of S0-2, this appears impossible\footnote{
If the central region is populated by a cluster of 
stellar-mass black holes as argued by \citet{bhcluster}, then the
local mass density is a factor $\sim 7$ lower, and hence the relaxation 
time is a factor $\sim 7$ longer.
}.

In this paper we propose that the orbit of SO-2 was a result
of the disruption of a massive-star binary 
in the tidal field of the massive black
hole at the Galactic center.  When the binary disrupts, one star is ejected
with energy greater than its incoming energy, 
leaving the other star, SO-2, in a bound orbit with the black hole.
In \S~\ref{sec:anal}, we
show analytically that for this mechanism to work, the former binary companion
of S0-2 must have been very massive, of order $100\,M_\odot$.
In \S~\ref{sec:numer}, we confirm this estimate by numerical simulation,
showing that the primary star must be $m_2\ga 60\,M_\odot$.
In \S~\ref{sec:discuss},
we discuss how the
detection of additional short-period ($P\la 100\,$yr) ``low''-eccentricity
($e\la 0.95$) companions to Sgr A* could be used to probe very high-mass
star formation in the inner pc of the Galactic center.
In Appendix \ref{asec:feed}, we discuss the mechanisms by which
such massive binaries might be fed to Sgr A*.  

\section{Analytic Estimates
\label{sec:anal}}

The specific binding energy of S0-2 today is
\begin{equation}
E_b = {GM_\bh(1-e)\over 2q}
\label{eqn:ebind}
\end{equation}
where $e=0.87$ and $q=125\,\au$ are the eccentricity and peribothron of
its orbit, and $M_\bh=3\times 10^6\,M_\odot$ is the mass of Sgr A*.  If S0-2
started out as part of a binary on a very low specific-energy 
(i.e., roughly parabolic) orbit with peribothron
$\sim q$,  then to become bound with $E_b$,
it would have to have been ejected from the binary with a
velocity
\begin{equation}
\Delta v = {E_b\over v} = \sqrt{G M_\bh\over 8 q}(1-e) \sim 210\,\kms,
\label{eqn:deltav}
\end{equation}
where $v=(2GM_\bh/q)^{1/2}$ is the velocity of the binary center of mass 
at peribothron.  If we now equate $\Delta v$ with the internal
velocity of S0-2 relative to the binary's center of mass,
\begin{equation}
\Delta v = \sqrt{G m_\tot\over a_\bin}{m_2\over m_\tot},
\label{eqn:deltav2}
\end{equation}
where $m_1$ and $m_2$ are the masses of S0-2 and its former companion,
$m_\tot=m_1+m_2$, and $a_\bin$ is the binary's semi-major axis, we obtain
\begin{equation}
a_\bin = q{8\over(1-e)^2}\,{m_2^2\over M_\bh m_\tot}.
\label{eqn:aeval}
\end{equation}
Finally, we assume that the binary can be broken apart tidally when the
accelerations are related by,
\begin{equation}
{2 G M_\bh a_\bin\over q^3} = \eta{G m_\tot\over a_\bin^2},
\label{eqn:eta}
\end{equation}
where $\eta\sim 1$ is a parameter.  We then find,
\begin{equation}
{m_2^3\over  m_\tot^2} = {\sqrt{\eta}\over 32}(1-e)^3\,M_\bh \sim
200\,\sqrt{\eta}\,M_\odot,
\label{eqn:meq}
\end{equation}
and
\begin{equation}
a_\bin{m_2\over m_\tot} = {\sqrt{\eta}\over 4} (1-e)\,\,q 
\sim 4\sqrt{\eta}\,\au.
\label{eqn:aeq}
\end{equation}
Equation (\ref{eqn:meq}) allows us to estimate the mass of the primary
star and equation (\ref{eqn:aeq}) the binary's semi-major
axis.
Note that in the limit, $m_2\gg m_1$, the left-hand sides 
of equations (\ref{eqn:meq}) and (\ref{eqn:aeq}) 
become $m_2$ and $a_\bin$, respectively.  On the other hand,
for a roughly equal mass ratio, $m_1\sim m_2\sim 800\eta^{1/2}\,M_\odot$. 
Such a massive S0-2 would be much brighter than is observed.   Hence,
we will generally work in the regime $m_2\gg m_1$.

Equation (\ref{eqn:meq}) implies that the former companion of S0-2
was quite massive,  with mass of order
several hundred solar masses.  However, since this equation contains
the parameter $\eta$, which depends on the specific internal orbital
parameters of the binary,  numerical simulations are required to 
determine the actual range of plausible masses for the companion.

\section{Numerical Estimates
\label{sec:numer}}

\subsection{Single Encounters
\label{sec:singenc}}

We conduct a series of simulations, in each of which the masses $m_1$
and $m_2$ are held fixed, but all of the other binary parameters are
allowed to vary.  The three binary orbital orientation angles and the
orbital phase are randomly chosen.  The binary eccentricity is drawn
uniformly in $e_\bin^2$ (but capped $e_\bin<0.95$), and the semi-major axis
is drawn uniformly in $\log a$ for $1\,\au<a<10\,\au$.  We find
that collisions leading to secondary orbits with the properties of SO-2's 
orbit, $q=125\,\au$ and $e=0.87$, have initial peribothra that are
about 5 AU larger, i.e., $q_\init\sim 130\,\au$.  All simulated collisions
therefore begin with the binary on a parabolic orbit with $q_\init$
at this value.  Each set of simulations contains 20,000 integrations
of the entire encounter between the massive-star binary and the massive 
black hole.

\begin{figure}
\plotone{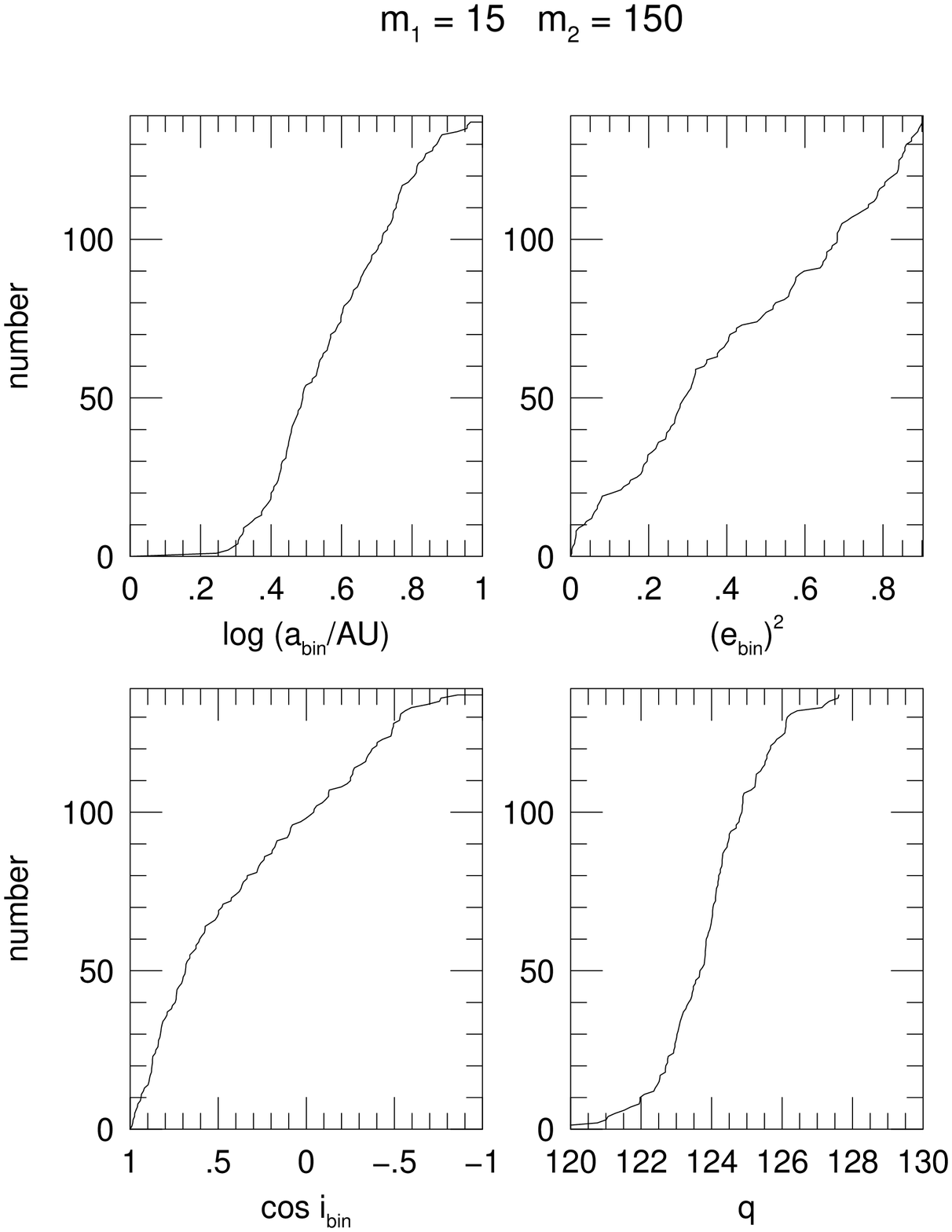}
\caption{\label{fig:cumdis}
Cumulative distributions for ``successful'' tidal collisions between a 
binary with masses $(m_1,m_2)=(15\,M_\odot,150\,M_\odot)$ and Sgr A*,
with ``success'' defined as $m_1$ being injected into an orbit with
eccentricity $0.86\leq e<0.88$ (similar to S0-2).  Shown are the
semi-major axis $a_\bin$, the eccentricity $e_\bin$ and the inclination
$i_\bin$ of the incoming binary, as well as the peribothron, $q$, of
the final $m_1$ orbit.  The simulated encounters were drawn uniformly
in $\log a_\bin$, $e_\bin^2$, and $\cos i_\bin$.  Hence ``success''
strongly favors prograde ($\cos i_\bin\sim 1$) orbits, and semi-major
axes $a_\bin\sim 4\,\au$, but does not depend strongly on $e_\bin$.
All simulated encounters had $q_\init=130\,\au$, while the final
peribothra are narrowly distributed around $125\,\au$.
}\end{figure}

For each set of simulations (determined by choice of $m_1$ and $m_2$)
we examine the cumulative distributions of three
binary input parameters ($a_\bin$, $e_\bin$, and $\cos i_\bin$) for the
subset of collisions that lead to an S0-2-like orbit, namely
$0.86\leq e < 0.88$.  We also examine the final peribothra $q$ of these
orbits to ensure that they are consistent with the observed value
for S0-2.  Figure \ref{fig:cumdis} shows the results for
the 137 collisions meeting this condition 
for the simulation with $m_1=15\,M_\odot$ and $m_2=150\,M_\odot$. 
Results for other mass combinations look similar. 

The first point to note from Figure 1 
is that less than 1\% of the 20,000 simulated
collisions lead to S0-2 type orbits.  Of course, this is partly because
we have defined the acceptable eccentricity range extremely narrowly.
However, many binaries simply do not break up, or if they do, they leave
S0-2 in a marginally bound or unbound orbit.  We return to this issue 
in \S~\ref{sec:repeatenc}

Of the parabolic collisions that do yield S0-2-like orbits, the distribution
of binary eccentricities is consistent with being uniform in $e_\bin^2$,
which is the same as the parent distribution of all collisions.  Hence,
the probability of producing an S0-2-like orbit does not strongly depend on
$e_\bin$.  By contrast, this probability depends strongly on the binary's
inclination $i$, with prograde orbits $(\cos i_\bin\sim 1)$ being much
preferred over retrograde orbits $(\cos i_\bin\sim -1)$.  This is plausible,
since it is easier to disrupt a prograde than retrograde binary.

As predicted by equation (\ref{eqn:aeq}), successful collisions have
$a_\bin\sim 4\,\au$.  There are essentially no successes for 
$a_\bin<2\,\au$.  The decline in success toward higher semi-major axis seems 
to be less severe than the decline toward lower values.  For this
reason, we have conducted additional simulations with input distributions
uniformly distributed of $1\,\au<a_\bin<100\,\au$.  We find that only
about 10\% of successful collisions have $a_\bin>10\,\au$, with the
overwhelming majority of these having $a_\bin<30\,\au$.

Finally, we find that the distribution of final peribothra is sharply peaked
near $q\sim 124\,\au$.  This shows that it is unnecessary to simulate
a broad range of $q_\init$ because only those binaries with $q_\init\sim 
130\,\au$ have the potential to yield S0-2-like orbits.

\begin{figure}
\plotone{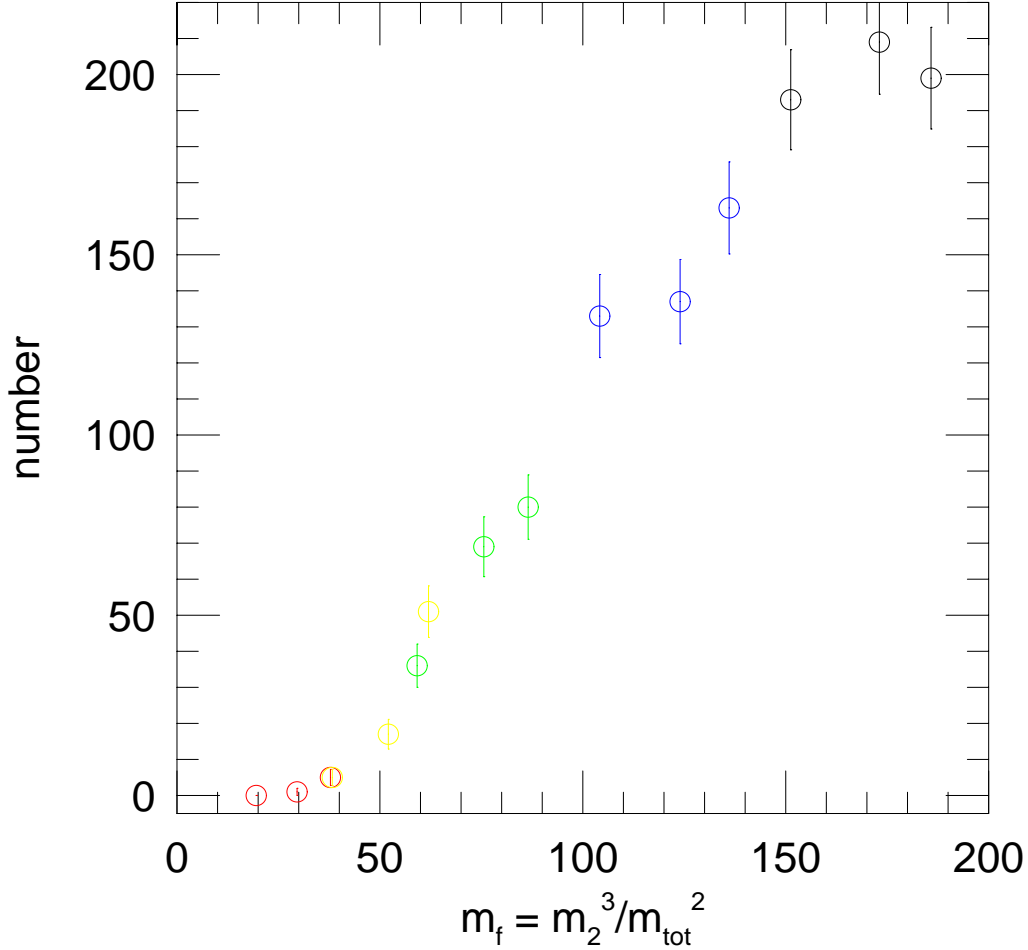}
\caption{\label{fig:massdis}
Number of ``successful'' tidal collisions (out of 20,000) between a 
binary with masses $(m_1,m_2)$ and Sgr A*, with ``success'' defined
as $m_1$ being injected into an orbit with eccentricity 
$0.86\leq e<0.88$ (similar to S0-2).  One may infer from eq.\ (\ref{eqn:meq}) 
that this success rate should be a function only of the combination
$m_f\equiv m_2^3/m_\tot^2$ with $m_\tot\equiv m_1+m_2$.  The figure bears
out this conjecture.  The red, yellow, green, blue, and black points 
refer respectively to $m_2=50,$ 75, 100, 150, and 200 $M_\odot$.  In each
case, trials were run for $m_1=7.5$, 15, and 30 $M_\odot$.  The success
rate peaks near $m_f\sim 200\,M_\odot$, also in agreement with the naive
prediction of eq.\ (\ref{eqn:meq}).
}\end{figure}

Figure \ref{fig:massdis} shows the number of successful collisions (again
defined by $0.86\leq e < 0.88$) as a function of the ``mass function''
$m_f\equiv m_2^3/m_\tot^2$ for 15 simulations with $m_1/M_\odot=7.5$, 15, 
and 30, and $m_2/M_\odot=50$, 75, 100, 150, and 200.  Note that, in
conformity with equation (\ref{eqn:meq}), the number of successful
collisions is approximately a function of $m_f$ rather than of the
two component masses separately. Also in conformity with
equation (\ref{eqn:meq}) is the fact that this function peaks at 
$m_f\sim 200\,M_\odot$.  There are essentially no successful
collisions for $m_f\la 40\,m_\odot$ which, for $m_1\sim 15\,M_\odot$,
corresponds to a minimum mass for the former companion of 
$m_2\ga 60\,M_\odot$.

As mentioned above, the absolute rate of successful collisions is
quite small.  Part of the problem is our extremely narrow definition
of ``success'': of course the a priori probability of exactly reproducing
the a posteriori known orbit of S0-2 is vanishingly small.  One might
plausibly argue that if S0-2 had any orbit with $q=125\,\au$ and
$e<0.9$ it might well have been measured.  Such broadened acceptance would
increase the number of ``successes'' by about a factor 4 near the peak.
The resulting fraction of successes is still fairly low.  

\subsection{Repeated Encounters
\label{sec:repeatenc}}

One should keep in mind, however, that
the orbital parameters of the incident binary can be significantly changed
even if it is not disrupted.  As discussed in Appendix \ref{asec:feed},
we expect that the incoming binary would not actually be 
on a parabolic orbit but rather on a highly
elliptical one with a semi-major axis $a\sim 10^4\,\au$, at which location 
a dense
young cluster would be tidally disrupted.  This induces only 
a slight ($\sim 5.4\%$) reduction in the energy requirements of the collision.
Within the formalism of \S~\ref{sec:anal}, the effect is equivalent to
decreasing $(1-e)$ by 5.4\%, or increasing $e$ from 0.87 to 0.877.  This
reduces the overall mass scale for $m_f=m_2^3/m_\tot^2$ by a factor
$(1-0.054)^3\sim 0.85$ (see eq. [\ref{eqn:meq}]).  Given the high mass scale 
of this equation and of Figure \ref{fig:massdis}, this reduction is welcome,
but modest.

However, this finite semi-major axis has another important effect:
the binary re-encounters Sgr A* on timescales of $\sim 10^3$ yrs.
When it does so, it preserves all the binary parameters it acquired
in the last collision, save the internal orbital phase relative to the
phase of the orbit around Sgr A*, which is effectively randomized.
Hence, thousands of such encounters may take place, which increase
the chance that the binary will be disrupted.  Partly, the internal binary
orbit changes each time.  However, even if the orbit is left unchanged, 
the mere randomizing of the internal 
orbital phase permits a new opportunity for a disruptive encounter.

To explore this effect, we simulate repeat encounters by binaries in orbit 
around Sgr A*, continuing the simulation until the binary is disrupted or
until it completes 1000 orbits.  In order to permit efficient computation,
we place the binaries in orbits with semi-major axis $a=3000\,\au$.  
Then, in order to be able to compare directly with the previous simulations
we consider as a success, collisions with the same energy change, that is,
with final orbits having eccentricity $e=0.847$.  Finally, to minimize the
role of Poisson fluctuations, we focus on the relative number successes
from the entire (up to 1000 orbit) simulation with the number coming
from the first encounter.  For $m_1=15\,M_\odot$ and $m_2=75\,M_\odot$, 
the 1000-orbit simulations doubles the number of successes, while for
$m_2=100\,M_\odot$ the improvement rises to about a factor 2.5.

About 90\% of the post-first-orbit successes occur within the first
100 orbits, and none past 200 orbits, indicating that our 1000-orbit
simulations were quite adequate.  The great majority of additional
successes have initial semi-major axes $1\,\au<a_\bin<2.5\,\au$.
Of those that do not, all but one are in retrograde orbits.  That is,
the additional successes arise because binaries with unfavorable parameters
are gradually perturbed into a more favorable regime.  Given the enhanced
probability for binaries with $a_\bin$ in the lowest range simulated,
we conduct an additional simulation with $m_2=100\,M_\odot$ and 
$0.1\,\au<a_\bin<1\,\au$.  However, we find very few successful collisions
from this range. 

In brief, allowing for multiple encounters in repeated orbits roughly
doubles the number of successful encounters, i.e., those leaving the secondary
in an orbit similar to SO-2.

\section{Discussion
\label{sec:discuss}}

	S0-2 is the only close companion of Sgr A* with well-determined
orbital parameters and a high quality spectroscopic identification.  
However, there could be many other remnants of
disrupted massive-star binaries still waiting to be discovered.
We found from our numerical integrations  that binaries were more often
disrupted leaving a star in a bound orbit with an eccentricity exceeding
that of SO-2.
We argued in \S~\ref{sec:numer} that the binary from which S0-2 was
ejected had a mass function of at least 
$m_f\equiv m_2^3/m_\tot^2\sim 40\,M_\odot$.  However, from
equation (\ref{eqn:meq}), binaries of this $m_f$ should typically eject stars
into orbits with eccentricity $e\sim 0.92$.  For orbits with the
same $q=125\,\au$ as S0-2, this would imply a period
longer by a factor $[(1-0.87)/(1-0.924)]^{3/2}\sim 2.2$, i.e., more than
30 years.  While such a star may have already been identified in near-infrared 
images, it would not yet be known whether the star would experience 
close approaches to Sgr A*.  
For example, the stars S1, S2 and S8 have estimated orbital
periods of around 2000 years.  However, their eccentricities are not well
constrained \citep{eckart}. 
Similarly, stars with the same eccentricity as
S0-2 but having larger peribothra would also have periods too long for one to
have yet determined their orbits.  Hence, there may well be a
substantial population of highly eccentric young stars orbiting Sgr A* 
that were injected into
their present orbits by the disruption of massive-star binaries.  If
so, these remnants could be used to probe statistically the recent formation of
very massive stars in the Galactocentric region, even though the massive
stars themselves have long since perished.

\acknowledgments 
We thank Jordi Miralda-Escud\'e and Dan Watson for stimulating discussions.
Work by AG was supported by grant AST 02-01266 from the NSF.
\bigskip
\appendix
\centerline{\bf APPENDIX}
\section{Feeding Delicacies to the Monster
\label{asec:feed}}

What mechanism might feed a substantial number of massive-star binaries 
to orbits with peribothra $q_\init\la 130\,\au$ from which Sgr A* can ``grab''
an S0-2-like star?  From \S~\ref{sec:numer}, only $\sim 5\%$ of binaries
with semi-major axes in the decade $1\,\au<a_\bin<10\,\au$ 
are devoured in this manner.  Moreover, only a small fraction, perhaps
$O(10\%)$, of massive stars will have a companion that is several times 
lighter orbiting with $a_\bin$ in this range.  Hence, perhaps only 
$\sim 0.5\%$ 
of massive stars passing within $q\sim 130\,\au$ will inject lighter
companions into S0-2-like orbits.  Since massive stars are fairly rare
and the phase space permitted for S0-2 progenitors is quite small, it would
seem necessary that these progenitors be preferentially funneled into
this phase space if the collisions discussed in \S\S~\ref{sec:anal} and 
\ref{sec:numer} are to plausibly account for the existence of S0-2.

\citet{gerhard}  suggested that the HeI stars in the central parsec
of the Galactic center were born in a $10^4-10^6M_\odot$
cluster that was formed more than 30pc away from Sag A* (like the Arches
and Quintuplet clusters).  The cluster sank inward to Sag A*
due to dynamical friction and was tidally disrupted by the massive black hole,
thus leaving massive stars within the central parsec.
Consider  a cluster of mass $M_\cl$ and velocity dispersion $\sigma_\cl$ that
has been formed outside the central pc of the Galactic center.
By the virial theorem, it will have an effective radius 
$R_\cl\sim G M_\cl/\sigma_\cl^2$.  The cluster will sink by dynamical
friction until it is tidally ripped apart at a separation from Sgr A*,
$r_d$,
\begin{equation}
r_d \sim R_\cl\biggl({M_\bh\over M_\cl}\biggr)^{1/3} \sim
{G M_\bh^{1/3}M_\cl^{2/3}\over \sigma_\cl^2}
= 1.7\times 10^4 \au
\biggl({M_\cl\over 10^4\,M_\odot}\biggr)^{2/3}
\biggl({\sigma_\cl\over 60 \,\kms}\biggr)^{-2}.
\label{eqn:acl}
\end{equation}
For a star from this disrupted cluster to come within $q_\init$ of Sgr A*,
it must have angular momentum $J<(2GM_\bh q_\init)^{1/2}$, and therefore
transverse velocity $v_\perp < (2GM_\bh q_\init)^{1/2}/r_d$.  Hence,
\begin{equation}
{v_\perp\over v_\circ} \sim {2\sigma_\cl\over v_q}\,
\biggl({M_\bh\over M_\cl}\biggr)^{1/3} = 0.12
\biggl({M_\cl\over 10^4M_\odot}\biggr)^{-1/3}
{\sigma_\cl\over 60\,\kms}
\label{eqn:vpvc}
\end{equation}
where $v_\circ=(G M/r_d)^{1/2}$ is the circular speed at breakup
and $v_q=6450\,\kms$ is the velocity at peribothron. Hence, if the
cluster is on a roughly circular orbit when it is disrupted, then essentially
none of its stars will have a close passage to Sgr A*.  

On the other hand, if the cluster is on a roughly radial orbit, then a 
fraction $f\sim 1 - \exp[(-q v_q/r_d\sigma_\cl)^2/2]$ will have close
passages.  That is,
\begin{equation}
f \sim -\ln(1-f)\sim 0.35
\biggl({M_\cl\over 10^4 M_\odot}\biggr)^{-4/3}
\biggl({\sigma_\cl\over 60\,\kms}\biggr)^{2},
\label{eqn:feval}
\end{equation}
so that for parameters that are plausible for a cluster formed in the
deep gravitational well of the Galactic center, a substantial fraction
of its members could come within $q_\init\sim 130\,\au$ 
following a radial-orbit
disruption.  In the above example, a massive-star cluster with 
$M_\cl\sim 10^4\,M_\odot$ would contain of order $10^2$ $100\,M_\odot$ stars.
Of these 1/3 would come within $q=130\,\au$ of Sgr A* and as estimated above,
of order 0.5\% of these might kick out companions onto S0-2-like orbits.
Over the lifetime of S0-2, several such clusters might form and
disrupt. Assuming that these clusters were on radial orbits, they
are a plausible source of S0-2-like stars.

The final question then is whether it is plausible that such clusters will
find themselves on radial orbits.  To address this question, we first 
show that for a power-law
density profile $\rho\propto r^{-\nu}$ in a background Kepler potential,
dynamical friction will tend to circularize orbits for $\nu>3/2$ and
make them more eccentric for $\nu<3/2$.  Since the maximum bound
velocity in a Kepler potential is $v_{\rm max}(r) = (2GM_\bh/r)^{1/2}$,
the amount of available phase-space scales as 
$[v_{\rm max}(r)]^3\propto r^{-3/2}$.  Hence, for $\nu=3/2$, the 
phase-space density $f({ u,r})$ is a constant, $f_0$, 
throughout the potential.  
The Chandrasekhar formula for the drag acceleration on a particle of
mass $m$ and velocity $\bf v$ is then ${\bf a}=-k{\bf v}$, where
\begin{equation}
k = {4\pi G^2 m_{\rm amb} m\over v^3}\int_0^v d u 4\pi u^2 f(u,r)\ln\Lambda
= {16\pi^2 G^2 m_{\rm amb}m f_0\over 3}\ln\Lambda,
\label{eqn:chandra}
\end{equation}
and where $m_{\rm amb}$ is the mean
mass of the ambient particles.  Since $\ln\Lambda$
depends only very weakly on radius, $k$ is essentially independent of radius.

Over the course of a single orbital period $P$, an object  with 
specific energy $E$ and specific angular momentum $\bf L$ will suffer 
mean rates of energy and angular momentum loss of
\begin{equation}
\langle{d E\over d t}\rangle = 
{\oint dt \,{\bf v \cdot a}\over P} = -{k\over P}\oint dt \,v^2 = 2kE
\label{eqn:eloss}
\end{equation}
and
\begin{equation}
\langle{d {\bf L}\over d t}\rangle = 
{\oint dt \,{\bf r \times a}\over P} = -{k\over P}\oint 
dt \,{\bf L} = -2k{\bf L}.
\label{eqn:lloss}
\end{equation}
Since $1-e^2= - 2L^2 E/(GM_\bh)^2$, the evolution of the
eccentricity can be written,
\begin{equation}
{d\ln(1-e^2)\over d t} = 2{\langle{d L/ d t}\rangle\over L} +
{\langle{d E/ d t}\rangle\over E} = -2k+2k=0.
\label{eqn:ecloss}
\end{equation}
For steeper profiles, the phase-space density is higher than average
at pericenter, hence the additional drag there tends to circularize
the orbit, while for shallower profiles the higher phase-space density at
apocenter tends to make them more eccentric.

Over most radii, the stellar mass profile around Sgr A* is a power law
with $\nu\sim 1.8$.  Hence, there is a weak tendency to circularize
orbits.  If there were a core in the mass distribution of radius $\sim 3''$,
corresponding to $\sim 2500\,\au$, then this core region could
drive clusters toward eccentric orbits.  That is, once the peribothron
fell below the core radius, the declining dynamical friction inside the
core (relative to that due to a power law) would drive the orbits
to greater eccentricity and so still lower peribothron.  

However, from equation (\ref{eqn:acl}), it is clear that this process could
not continue very long before the cluster was disrupted, unless the
cluster had rather extreme parameters.  On the other hand, from
equation (\ref{eqn:vpvc}), the cluster orbit must be highly eccentric
if a significant fractions of its disrupted contents is to reach
$q_\init$.  Dynamical friction by itself probably cannot produce such
orbits.  If a cluster had the required orbit to deliver 
S0-2 progenitors to Sgr A*, it must have been born on it.  This in itself
is not implausible: it is certainly possible for clusters to form from
colliding clouds of gas whose angular momenta roughly cancel.  Whether
this happens frequently or infrequently is today, however, a matter of
speculation.


\clearpage

\begin{thebibliography}{}

\bibitem[Alexander(1999)]{alexander} Alexander, T.\ 1999, \apj, 527, 835

\bibitem[Eckart et al.(2002)]{eckart} Eckart, A., Genzel, R., Ott, T.,  \&
Schodel, R.\ 2002, MNRAS, 331, 917 

\bibitem[Eckart \& Genzel(1996)]{eckgen} Eckart, A. \& Genzel, R.\ 1996, 
Nature, 383, 415

\bibitem[Ghez et al.(1998)]{ghez98} Ghez, A., Klein, B.L., Morris, M.,
\& Becklin, E.E.\ 1998, \apj, 509,678

\bibitem[Ghez et al.(2003)]{ghez03} Ghez, A., et al.\ 2003, \apj 
\ Letters, in press (astro-ph/0302299)

\bibitem[Gerhard(2001)]{gerhard}Gerhard, O.\ 2001, \apj, 546, L39

\bibitem[Miralda-Escud\'e \& Gould(2000)]{bhcluster} Miralda-Escud\'e., 
\& Gould, A.\ 2000, \apj, 545, 847

\bibitem[Salim \& Gould(1999)]{visbin} Salim, S., \& Gould, A.\ 1999,
\apj, 523, 633

\bibitem[Schodel et al.(2002)]{schodel} Schodel, R., et al.\ 2002,
Nature, 419, 694


\end{thebibliography}
\end{document}